\documentclass{optica-article}

\journal{opticajournal} % for journals or Optica Open

\articletype{Research Article}

\usepackage{lineno}
\begin{document}

\title{White-Light Matter-Wave Interferometry} %Enables Ambiguity-Free Inertial Sensing}
%\title{Multi-Wavelength Atom Interferometry for Inertial Measurement}

%\author{Wei-Chen Jia,\authormark{1,2} Yue-Xin,\authormark{1,2,*} Ke Shen and Author Three\authormark{,*}}

\author{Wei-Chen Jia, Yue-Xin, Ke Shen, Zhi-Xin Meng and Yan-Ying Feng\authormark{*}}

\address{\authormark{1}A-Knows Lab, Department of Precision Instruments, Tsinghua University, Beijing 100084, China\\
\authormark{2}State Key Laboratory of Precision Measurement Technology and Instruments, Beijing 100084, China\\
}

\email{\authormark{*}yyfeng@tsinghua.edu.cn} %% email address is required; see note below about the corresponding author designation

% use {asbstract*} to suppress the copyright line. Copyright information will be added in production

\begin{abstract*} 
White-light interferometry exploits broadband optical spectra to generate a localized interference envelope, providing a unique reference beyond the periodicity of individual fringes and underpinning numerous applications in precision metrology, ranging, and imaging. Despite the common wave nature of light and matter, a matter-wave counterpart of white-light interferometry has remained unexplored. Here we report the first realization of white-light matter-wave interferometry, extending this established optical paradigm to matter waves. Using a continuous atomic beam as a broadband matter-wave source, we synthesize velocity-dependent spectral components to generate a localized matter-wave interference envelope, introducing a new measurement observable based on envelope localization rather than conventional fringe-phase estimation. As a proof-of-principle demonstration, we implement simultaneous dual-axis rotation and acceleration sensing with a white-light Mach--Zehnder atom interferometer, revealing three defining features: envelope localization beyond phase ambiguity, a velocity-engineered measurement response, and intrinsic rejection of phase bias. We further measure the Earth's rotation with a relative error of 4.3$\%$ and achieve a long-term stability of 93 ppm relative to the Earth's rotation rate after 15,000 s of averaging. These results establish a new interferometric mechanism in which ambiguity, measurement response, and robustness can be addressed through the intrinsic design of the broadband matter-wave interference itself, extending the principles of optical white-light interferometry to atom optics and opening new opportunities for quantum sensing, precision metrology, and matter-wave metrology.

\bigskip

\end{abstract*}

%%%%%%%%%%%%%%%%%%%%%%%%%%  body  %%%%%%%%%%%%%%%%%%%%%%%%%%
\section{Introduction}

Interference provides one of the most powerful mechanisms for precision measurement. However, measurements based on periodic interference fringes inherently face a trade-off between sensitivity and unambiguous range: fine fringes provide high resolution but introduce ambiguity when the phase excursion exceeds a single fringe. This limitation is particularly relevant to atom interferometers, where it manifests as phase ambiguity, restricted dynamic range, and sensitivity to unknown phase offsets \cite{bongs2019taking,narducci2022advances}.

Optical white-light interferometry overcomes this limitation by exploiting spectral bandwidth. The superposition of broadband optical components produces a localized coherence envelope that provides a unique positional reference, while the underlying high-frequency fringes retain high measurement precision \cite{al1983partially,Hariharan1994,harasaki2000fringe,gomez2020noise}. This combination of global localization and fine interferometric resolution has made white-light interferometry a powerful tool in optical metrology, including ranging, surface profiling, and dimensional measurement. Matter waves possess an analogous spectral degree of freedom: an atomic beam naturally contains a continuous distribution of velocities, and hence a broadband distribution of de Broglie wavelengths. Despite this direct correspondence between optical and matter-wave spectra, a matter-wave counterpart of white-light interferometry has remained unexplored.

Here we introduce and experimentally realize white-light matter-wave interferometry, establishing the matter-wave counterpart of optical white-light interferometry. In this correspondence, the broadband optical spectrum is replaced by the intrinsic velocity spectrum of an atomic beam, while distinct optical spectral components are mapped onto different matter-wave velocity classes. Their synthesis produces a localized matter-wave interference envelope, directly mirroring the formation of the coherence envelope in optical white-light interferometry. The envelope therefore provides a global localization reference beyond the periodicity of individual matter-wave fringes and introduces a new interferometric observable based on envelope position rather than exclusively on fringe phase. More broadly, the spectral degree of freedom provides an additional handle for engineering the response and robustness of matter-wave interferometry.

We demonstrate this new interferometric regime using a white-light Raman Mach--Zehnder atom interferometer, with inertial sensing serving as a proof-of-principle application. Three defining features are experimentally observed: envelope localization beyond phase ambiguity, a velocity-engineered measurement response manifested here as a velocity-independent rotational scale factor, and intrinsic rejection of phase bias. These features are verified through simultaneous rotation and acceleration measurements, followed by a long-term measurement of the Earth's rotation that demonstrates accurate and robust absolute inertial readout. Unlike existing approaches that extend the operating range of atom interferometers through classical-sensor hybridization \cite{lautier2014hybridizing,merlet2009operating,cheiney2018navigation}, scale-factor or multi-configuration reconstruction \cite{bonnin2018a,avinadav2020,yankelev2020a,black2023velocity,chen2024self}, or closed-loop operation \cite{d2024atom,sato2025closed,jia2026closed}, white-light matter-wave interferometry introduces localization directly into the interferometric signal through broadband synthesis. Although demonstrated here through inertial sensing, the underlying concept is more general, extending the established paradigm of optical white-light interferometry into the matter-wave domain, opening a previously unexplored regime of matter-wave interferometry in which broadband spectral synthesis can be used to create and engineer new interferometric observables.

\section{Principle of White-Light Matter-Wave Interferometry}

Light-pulse atom interferometry provides a versatile platform for coherent matter-wave manipulation and precision measurement\cite{hamiltonDark2015a,rosi2014precision,zhou2015test,durfee2006long,gautierAccurate2022}. Laser pulses act as coherent matter-wave beam splitters, mirrors, and phase references that define the interferometric ruler. As atomic wave packets sequentially interact with the light pulses, they accumulate the optical phases imprinted at each interaction, giving rise to a well-defined interferometric phase. Physical quantities of interest couple to the interferometric evolution by modifying this accumulated phase, producing an interferometric phase shift that serves as the fundamental measurement observable.

\begin{figure}
    \centering
    \includegraphics[width=1\linewidth]{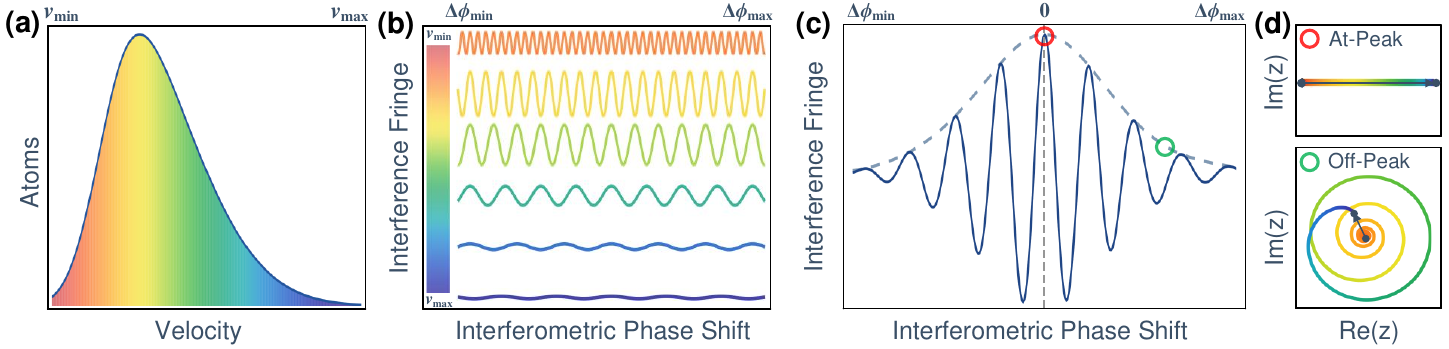}
    \caption{White-light matter-wave interferometry.
(a) Longitudinal velocity distribution of the atomic beam source, forming a broadband matter-wave spectrum.
(b) Raman Mach--Zehnder interference fringes generated by different velocity classes.
(c) Synthesis of the broadband matter-wave spectrum, giving rise to a localized matter-wave interference envelope. The envelope peak defines the unique zero-reference of the induced interferometric phase shift.
(d) Phasor representation of the contributions from different velocity classes at the envelope peak (red circle) and off-peak (green circle) positions indicated in (c).
}
\label{fig:Schematic}
\end{figure}

Unlike conventional atom interferometers that operate effectively with a single characteristic scale factor, white-light matter-wave interferometry exploits the intrinsic velocity bandwidth of an atomic beam as a broadband matter-wave spectrum. As shown in Fig.~\ref{fig:Schematic}(a), the transversely cooled atomic beam possesses a continuous longitudinal velocity distribution. Each velocity class corresponds to a distinct de Broglie wavelength, giving rise to a different interferometric phase evolution. The atomic ensemble therefore constitutes a broadband matter-wave source, establishing the physical basis of white-light matter-wave interferometry.

As illustrated in Fig.~\ref{fig:Schematic}(b), for a physical quantity $x$, atoms belonging to a velocity class $v_i$ generate a sinusoidal interferometric response,

\begin{equation}
S(v_i)=
A(v_i)+\frac{C(v_i)}{2}
\cos\!\left[
\Delta\phi(v_i,x)+\phi_0
\right],
\label{eq:singlev}
\end{equation}
where $A(v_i)$ and $C(v_i)$ denote the signal offset and contrast, respectively, and $\phi_0$ is the initial interferometric phase. The quantity
\begin{equation}
\Delta\phi(v_i,x)=f(v_i)x
\label{eq:phase_shift}
\end{equation}
represents the interferometric phase shift induced by the physical quantity $x$, where $f(v_i)$ is the velocity-dependent scale factor associated with the velocity class $v_i$. Equation~(\ref{eq:singlev}) shows that each velocity class constitutes an individual interferometer with a distinct scale factor, giving rise to interference fringes with different periodicities as the interferometric phase shift is varied. 

The detected interferometric signal arises from the ensemble-averaged contributions of different atomic velocity classes. The resulting broadband response can be described by introducing a complex representation of the interference contributions,

\begin{equation}
    \mathcal{P}(x)=\int_{v_{\min}}^{v_{\max}} W(v)e^{i\Phi(v,x)}\,dv,
\end{equation}
where $\Phi(v,x)=\Delta\phi(v,x)+\phi_0$ is the total interferometric phase, and $W(v)=\rho(v)C(v)\ge0$ denotes the velocity-resolved interference-amplitude density, incorporating the atomic velocity distribution $\rho(v)$ and the corresponding interferometric contrast $C(v)$. The measured signal is given by $S_{\mathrm{tot}}(x)=A_{\mathrm{tot}}+\operatorname{Re}[\mathcal{P}(x)]/2$, where $S_{\mathrm{tot}}$ and $A_{\mathrm{tot}}$ represent the ensemble-averaged detected signal and its background, respectively.

The interference contribution from each velocity class can be represented by a phasor whose magnitude is determined by $W(v)$ and whose orientation is determined by $\Phi(v,x)$. The resultant complex amplitude $\mathcal{P}(x)$ is obtained by summing these phasors, and its magnitude,
\begin{equation}
\mathcal{E}(x)=|\mathcal{P}(x)|,
\end{equation}
defines the broadband matter-wave interference envelope. 

The physical origin of envelope localization lies in the velocity-dependent phase dispersion. At $x=0$, the induced phase shift $\Delta\phi(v,x)$ vanishes for all velocity classes, causing their phasors to align and produce the maximum resultant amplitude [upper panel of Fig. 1(d)]. This condition can be rigorously established through
\begin{equation}
    \mathcal{E}(x)=\left|\int_{v_{\min}}^{v_{\max}}W(v)e^{i\Phi(v,x)}dv\right|
\leq\int_{v_{\min}}^{v_{\max}}W(v)dv=\mathcal{E}(0),
\end{equation}
Equality holds when all non-zero contributions to the integral share the same complex argument, corresponding to complete phase alignment across the velocity spectrum. As $x$ deviates from zero, the different scale factors lead to increasing phase dispersion across the velocity spectrum. The phasors progressively lose their alignment and partially cancel one another [lower panel of Fig. 1(d)], reducing the resultant amplitude and giving rise to a localized interference envelope, as shown in Fig. 1(c). 

Crucially, the envelope maximum identifies the common zero-reference of the induced interferometric phase shift, where all velocity classes are simultaneously phase-aligned. The velocity-independent phase offset $\phi_0$ rotates all phasors by the same angle without changing their resultant magnitude. Consequently, the envelope peak position is determined by the relative phase evolution across the velocity spectrum rather than by the absolute interferometric phase.

White-light matter-wave interferometry therefore introduces a fundamentally different measurement paradigm for matter waves, which exploits the position of a localized broadband interference envelope as a new interferometric observable. By establishing a unique reference through the intrinsic velocity-dependent phase evolution of matter waves, this approach enables ambiguity-free measurement without relying on conventional fringe-phase tracking. In the following, we employ atom-interferometric inertial sensing as a proof-of-principle application to illustrate and experimentally validate this measurement principle.

\section{Implementation of white-light matter-wave interferometry for inertial sensing}

Atom-interferometric inertial sensing is one of the most important applications of light-pulse atom interferometry, with broad impact in navigation \cite{durfee2006long,canuel2006six,Meng2024, Yan2025,sato2025closed}, geophysics \cite{stockton2011absolute,rosi2014precision, gautierAccurate2022,stray2022quantum}, and precision metrology \cite{tino2007atom,graham2016dark,abe2021matter}. As a proof-of-principle demonstration of white-light matter-wave interferometry, we implement a white-light Raman Mach--Zehnder (Raman M--Z) atom interferometer and employ inertial sensing as a representative application. This implementation allows us to experimentally validate the operating principle of envelope-based interferometric readout and to demonstrate its advantages over conventional phase-based measurements.

\begin{figure}
    \centering
    \includegraphics[width=1\linewidth]{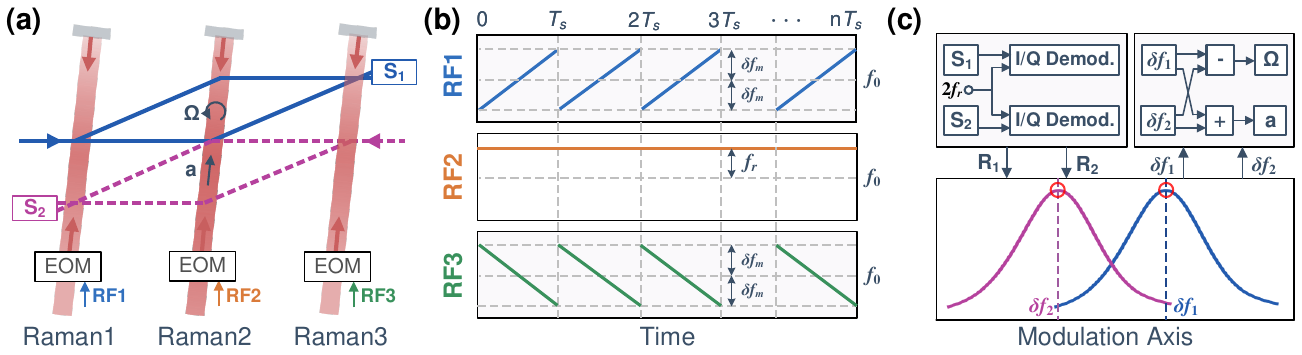}
    \caption{
Experimental implementation of white-light matter-wave interferometric inertial sensing.
(a) Experimental implementation of the dual-axis white-light Raman M--Z atom interferometer.
(b) Controlled modulation of the Raman two-photon detunings for reconstruction of the white-light interference envelopes.
(c) Envelope reconstruction through quadrature demodulation. The peak positions of the two interference envelopes, $\delta f_1$ and $\delta f_2$, are extracted to determine the rotation and acceleration through differential and common-mode processing, respectively.
}
\label{fig:Setup}
\end{figure}

Figure~\ref{fig:Setup}(a) illustrates the experimental implementation of the proposed white-light matter-wave interferometry based on a Raman M--Z atom interferometer. Transversely cooled $^{87}$Rb atomic beams provide the broadband matter-wave spectrum required for white-light matter-wave interferometry. Two-photon stimulated Raman transitions coherently manipulate the atomic matter waves by successively acting as beam splitters, mirrors, and recombiners, thereby forming a closed interferometric loop. The physical quantity to be measured is encoded as an interferometric phase shift accumulated during this evolution. For inertial sensing, the interferometer is simultaneously sensitive to rotations $\Omega$ about the normal of the enclosed interferometer area and accelerations $a$ along the effective Raman wave vector. To enable simultaneous dual-axis inertial sensing, two counter-propagating atomic beams are employed to construct a pair of Raman M--Z interferometers, allowing rotation and acceleration to be naturally decoupled through the average and difference of the two interferometric phases.

The interferometric signal for atoms with longitudinal velocity $v_i$ in a Raman M–Z interferometer is given by
\begin{equation}
    S(v_i)=A(v_i)+\frac{C(v_i)}{2}cos(k_{\mathrm{eff}} a \frac{L^2}{v_i^2}- 2 k_{\mathrm{eff}} \Omega \frac{L^2}{v_i}+ \phi_0),
\end{equation}
where $k_{\mathrm{eff}}$ is the effective Raman wave vector, $L$ is the spacing between adjacent Raman interaction zones, and $\phi_0$ represents a non-inertial phase contribution. In this specific implementation, the generalized velocity-dependent phase shift introduced in the preceding section is realized by the acceleration- and rotation-induced terms, which scale as $v_i^{-2}$ and $v_i^{-1}$, respectively. Consequently, different velocity classes produce interference fringes with distinct inertial phases, and their superposition forms the white-light interference envelope. Following the general principle established above, the envelope reaches its maximum when the velocity-dependent inertial phase shifts are aligned, with its peak providing a unique reference for the zero inertial phase-shift condition. 

This envelope-based interferometric response provides a new route for ambiguity-free inertial sensing beyond conventional single-phase readout. However, direct reconstruction of the interference envelope would, in principle, require scanning the inertial quantity (i.e., the induced inertial phase shift) itself, which is incompatible with practical sensing applications. We therefore modulate the interferometric phase to generate a controllable phase shift equivalent to the inertial phase shift, allowing the white-light envelope to be reconstructed and localized without scanning the inertial input itself.

As shown in Fig.~\ref{fig:Setup}(a), the Raman lasers are independently modulated using electro-optic modulators (EOMs), generating dual frequency components that drive two-photon stimulated Raman transitions between the hyperfine states $^{87}$Rb $\lvert 5^2\mathrm{S}_{1/2},F=1 \rangle$ and $^{87}$Rb $\lvert 5^2\mathrm{S}_{1/2},F=2 \rangle$. The frequency difference is given by 
$\Delta_i=2\pi(f_0+\delta_i)\,(i=1,2,3)$,
where $f_0$ includes the hyperfine splitting together with Doppler and recoil shifts, and $\delta_i$ denotes a controllable two-photon detuning. By tuning $\delta_i$, an additional phase shift is introduced into the interferometer, effectively scanning the interferometric phase. This modulation is formally   equivalent to probing different inertial phase values and therefore enables reconstruction of the white-light matter-wave envelope.

As shown in Fig.~\ref{fig:Setup}(b), the detunings of Raman lasers 1 and 3 (i.e., the two $\pi/2$ pulses) are synchronously scanned in opposite directions with a period $T_s$, while Raman laser 2 (i.e., the $\pi$ pulse) is applied with a constant detuning $f_r$. The resulting interferometric phases become
\begin{equation}
\begin{aligned}
\Phi_1' =k_{\mathrm{eff}} a T^2- 2 k_{\mathrm{eff}} \Omega L T- 4\pi f_r (t_0 + T)+ 4\pi \delta f T+ \phi_0,
\\
\Phi_2'=- k_{\mathrm{eff}} a T^2- 2 k_{\mathrm{eff}} \Omega L T- 4\pi f_r (t_0 + T)- 4\pi \delta f T+ \phi_0'.\label{eq:mod-phi}
\end{aligned}
\end{equation}
where $T=L/v$ denotes the atom transit time between adjacent Raman interaction zones. 

Quadrature demodulation is then applied using a reference signal at frequency $2f_r$. This procedure removes the rapidly varying term $4\pi f_r t_0$ and yields orthogonal components $X_i$ and $Y_i\,(i=1,2)$, from which the envelope amplitudes are obtained as $R_i = \sqrt{{X_i}^2 + {Y_i}^2}$. The corresponding phases, $\Phi_i=\arctan({X_i}/{Y_i})$, are  
\begin{equation}
\begin{aligned}
\Phi_1 =k_{\mathrm{eff}} a T^2- 2 k_{\mathrm{eff}} \Omega L T- 4\pi f_r  T+ 4\pi \delta f T+ \phi_0,
\\
\Phi_2=- k_{\mathrm{eff}} a T^2- 2 k_{\mathrm{eff}} \Omega L T- 4\pi f_r  T- 4\pi \delta f T+ \phi_0'.
\label{eq:mod-phi1}
\end{aligned}
\end{equation}

According to the phase-alignment criterion established in the previous section, the envelope peak is reached when the phase dispersion across the broadband matter-wave spectrum vanishes, namely

\begin{equation}
\frac{\partial \Phi_i}{\partial v}=0.
\label{eq:partial}
\end{equation}

This condition yields
\begin{equation}
\begin{aligned}
\delta f_1 &=
\frac{k_{\mathrm{eff}}L}{2\pi}\Omega
-\frac{k_{\mathrm{eff}}L}{2\pi v_{\mathrm{eq}}}a
+f_r,
\\
\delta f_2 &=
-\frac{k_{\mathrm{eff}}L}{2\pi}\Omega
-\frac{k_{\mathrm{eff}}L}{2\pi v_{\mathrm{eq}}}a
-f_r,
\end{aligned}
\end{equation}
Here, for acceleration sensing, the different velocity dependences of the inertial and modulation phases prevent exact phase alignment across the entire ensemble. The envelope maximum is therefore characterized by an effective ensemble velocity $v_{\mathrm{eq}}$.

By summing and differencing the two envelope peak positions, the rotation and acceleration are directly decoupled,

\begin{equation}
\begin{aligned}
\delta f_1-\delta f_2
&=
\frac{k_{\mathrm{eff}}L}{\pi}\Omega
+2f_r,
\\
\delta f_1+\delta f_2
&=
-\frac{k_{\mathrm{eff}}L}{\pi v_{\mathrm{eq}}}a.
\end{aligned}
\label{eq:MW-mea}
\end{equation}

This decoupling can be understood more intuitively from the motion of the two white-light interference envelopes. An acceleration input translates the two envelopes rigidly along the detuning axis in the same direction, leaving their separation unchanged while shifting their midpoint. In contrast, a rotation input drives the two envelopes toward or away from each other, changing their separation while keeping the midpoint fixed. Consequently, the midpoint of the two envelope peaks provides a direct measure of acceleration, whereas their separation uniquely determines rotation.

Compared with conventional atom interferometers based on fringe/phase-only readout, white-light matter-wave interferometry provides several fundamental advantages:

(i) \textbf{Envelope localization beyond phase ambiguity}—
The interference envelope provides a unique global reference that enables ambiguity-free inertial estimation beyond the conventional half-fringe limit. Unlike standard atom interferometers, whose outputs repeat periodically with phase, the envelope peak remains uniquely identifiable over a substantially extended measurement range.

(ii) \textbf{Velocity-engineered measurement response}—
The modulation-induced phase and the rotation-induced phase exhibit identical velocity dependence, resulting in a rotational scale factor that is intrinsically independent of atomic velocity. This suppresses sensitivity to velocity drifts and improves long-term stability. More generally, by engineering the velocity dependence of different phase contributions, the dependence of the measured quantity on atomic velocity can be deliberately tailored, providing a new degree of freedom for matter-wave interferometric sensing.

(iii) \textbf{Intrinsic rejection of phase bias}—
The envelope peak position depends on phase alignment across the broadband matter-wave spectrum rather than on the absolute interferometric phase itself. Consequently, velocity-independent phase biases contribute weakly to the inferred inertial signal, improving robustness against systematic errors.

Overall, the proposed approach replaces conventional phase-only interferometric readout with envelope localization based on a broadband matter-wave spectrum. By extending the principles of white-light interferometry from optics to matter waves, it establishes a new interferometric paradigm for ambiguity-free, wide-range, and high-precision inertial sensing.

\section{Experimental Realization}

In the implemented white-light atomic interferometric inertial sensor, a pair of counter-propagating $^{87}$Rb atomic beams are generated from dual atomic ovens and transversely collimated by two-dimensional optical molasses (2D OM) formed by cooling lasers. The resulting atomic beams exhibit a transverse temperature of $524\,\mu\mathrm{K}$, a most probable longitudinal velocity of approximately $190\,\mathrm{m/s}$, and a full width at half maximum (FWHM) of $320\,\mathrm{m/s}$. This broad velocity distribution forms a broadband matter-wave spectrum that serves as the source for white-light matter-wave interferometry.

After state preparation, the two atomic beams interact with a common sequence of three Raman laser pulses in a $\pi/2$--$\pi$--$\pi/2$ Mach--Zehnder configuration. The Raman interaction zones are separated by a distance of $2L=540~\mathrm{mm}$, and the Raman beams have a waist of $1~\mathrm{mm}$. Following the interferometric sequence, atoms in the $\lvert 5^2\mathrm{S}_{1/2},F=2\rangle$ state are detected via laser-induced fluorescence, yielding two interferometric signals, $S_1$ and $S_2$ \cite{Yan2025}.

Envelope reconstruction is realized through controlled phase modulation of the Raman pulses. The required two-photon detunings are generated by radio-frequency signals referenced to a low-phase-noise microwave source, enabling independent control of the detuning applied to each Raman pulse. Specifically, the detunings of the first and third Raman pulses are synchronously scanned in opposite directions over the range $(-\delta f_m,+\delta f_m)$, while the second pulse is maintained at a constant detuning of $f_r=200~\mathrm{Hz}$. This modulation scheme realizes the effective phase-space scan described in the previous section and enables localization of the interference envelope without varying the physical inertial input.

The interference signals are processed using quadrature demodulation with a reference frequency of $f_{\mathrm{ref}}=2f_r$. After low-pass filtering with a cutoff frequency of $20~\mathrm{Hz}$, the envelope amplitudes $R_1$ and $R_2$ are extracted. The corresponding envelope peak positions are determined through Gaussian fitting. To reduce noise-induced bias, the data are first smoothed and a local region surrounding the envelope maximum is selected for fitting. The peak locations, denoted by $\delta f_1$ and $\delta f_2$, are subsequently obtained from the fitted Gaussian profiles.

Dual-axis inertial calibration is performed using a high-precision air-bearing rotation stage and a single-axis tilt platform. Rotation measurements are realized by applying calibrated angular velocities about the sensitive axis, while acceleration measurements are performed by varying the projection of gravity along the effective Raman wave vector.

Together, these implementations provide the experimental platform for the proof-of-principle validation of white-light matter-wave interferometry presented in the following sections.

\section{Results and Discussion}

\begin{figure*}
\centering
\includegraphics[width=1\linewidth]{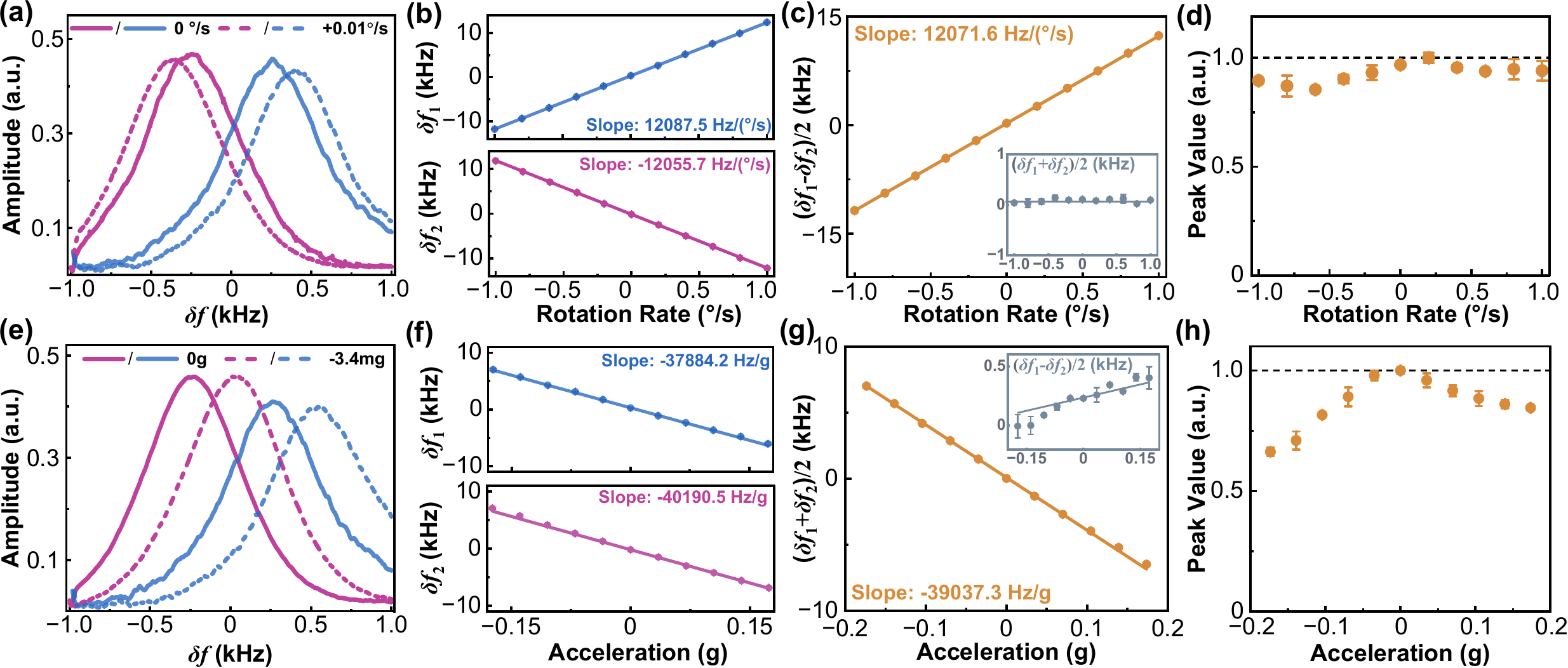}
\caption{Experimental demonstration of envelope localization and velocity-engineered measurement response in white-light matter-wave interferometric inertial sensing.
(a) Measured matter-wave interference envelopes at different rotation rates.
(b) Individual envelope peak positions as a function of rotation rate.
(c) Rotation measurement based on half of the envelope separation as a function of rotation rate. Inset: mean envelope peak position versus rotation rate, demonstrating negligible rotation-to-acceleration cross-coupling.
(d) Normalized envelope peak value versus rotation rate.
(e) Measured matter-wave interference envelopes at different acceleration inputs.
(f) Individual envelope peak positions as a function of acceleration.
(g) Acceleration measurement based on the mean envelope peak position as a function of acceleration. Inset: half of the envelope separation versus acceleration, revealing residual acceleration-to-rotation cross-coupling.
(h) Normalized envelope peak value versus acceleration.
}
\label{fig:inertial}
\end{figure*}

\subsection{Feature I: Envelope localization beyond phase ambiguity}

The first defining feature of white-light matter-wave interferometry is the transformation of the measurement observable from a periodic fringe phase to a localized interference envelope. %By synthesizing a broadband matter-wave spectrum, white-light matter-wave interferometry introduces an additional localization scale: the envelope peak. The measured quantity can therefore be identified from the envelope position rather than solely from the phase of an individual fringe, providing a global reference beyond the intrinsic phase ambiguity of conventional interferometry.

We experimentally demonstrate this feature through inertial sensing. Figure~\ref{fig:inertial}(a) shows the measured interference envelopes under different rotation rates. As predicted by Eq.~(\ref{eq:MW-mea}), the two envelopes translate in opposite directions along the modulation axis $\delta f$ while maintaining nearly identical shapes. Their peak positions therefore remain well localized over the entire investigated rotation range.

The evolution of the individual envelope peaks is quantified in Fig.~\ref{fig:inertial}(b). The two peak positions exhibit equal slopes with opposite signs, such that half of their separation provides the rotation observable [Fig.~\ref{fig:inertial}(c)]. A scale factor of $12071.6\,\rm{Hz/(^\circ/s)}$ is obtained, in excellent agreement ($0.06\%$ deviation) with the theoretical prediction of $\frac{k_{\mathrm{eff}}L}{2\pi}$. The small discrepancy is attributed to calibration uncertainties of the rotation stage. Throughout the investigated range of $\pm1^\circ/\mathrm{s}$, the normalized envelope peak value remains nearly unchanged [Fig.~\ref{fig:inertial}(d)], demonstrating robust envelope localization far beyond the conventional half-fringe limit. Compared with the unambiguous range of approximately $\pm\frac{\pi/2}{2k_{\mathrm{eff}}LT}\approx0.007^\circ/\mathrm{s}$, the envelope-based measurement extends the accessible rotation range by approximately two orders of magnitude.

The same envelope-localization principle is verified for acceleration measurements. As shown in Figs.~\ref{fig:inertial}(e,f), the two envelopes translate in the same direction along the $\delta f$ axis, and their mean peak position provides the acceleration observable [Fig.~\ref{fig:inertial}(g)]. Unlike the rotational response, the normalized envelope peak value gradually decreases with increasing acceleration [Fig.~\ref{fig:inertial}(h)]. This reduction originates from the different velocity dependences of the modulation phase $4\pi\delta fT$ and the acceleration-induced phase $k_{\mathrm{eff}}aT^2$, which introduce residual higher-order phase dispersion beyond the stationary-phase condition \cite{jia2026closed}. Nevertheless, the envelope peak remains clearly localized throughout the investigated range of $\pm0.17\,g$, corresponding to the maximum tilt angle ($\pm10^\circ$) of the calibration platform. Compared with the conventional half-fringe range of approximately $\pm\frac{\pi/2}{k_{\mathrm{eff}}T^2}\approx0.005\,g$, white-light matter-wave interferometry extends the ambiguity-free acceleration range by approximately a factor of 34.

More broadly, envelope localization represents a fundamental change in how information can be encoded and retrieved in matter-wave interferometry. Rather than relying exclusively on the periodic phase of a single interference fringe, broadband synthesis introduces a non-periodic localization observable that complements the underlying high-resolution interference structure. White-light matter-wave interferometry therefore provides a general route for overcoming phase ambiguity through the design of the interferometric observable itself, rather than through auxiliary phase reconstruction or external reference information.

\subsection{Feature II: Velocity-engineered measurement response}

The second defining feature of white-light matter-wave interferometry is the ability to engineer measurement responses. Because different physical quantities generally exhibit different velocity scaling, appropriately designed phase modulation can selectively match or compensate these velocity dependences, thereby tailoring the measurement response beyond that of a conventional single-wavelength interferometer.

A representative example is the rotational response demonstrated here. Since the modulation-induced phase and the rotation-induced phase share the same velocity dependence ($\propto v^{-1}$), the envelope displacement becomes intrinsically independent of atomic velocity. By contrast, the acceleration-induced phase scales as $v^{-2}$, leading to a residual dependence on the effective velocity of the contributing atomic ensemble. This difference results in fundamentally different coupling behaviors for the two sensing channels.

The individual responses of the two envelopes versus acceleration input are shown in Fig.~\ref{fig:inertial}(f), corresponding to effective velocities of $v_1=179\,\mathrm{m/s}$ and $v_2=169\,\mathrm{m/s}$, respectively. The resulting velocity mismatch gives:

\begin{equation}
\begin{aligned}
\delta f_1-\delta f_2 &=
\frac{k_{\mathrm{eff}}L}{\pi}\Omega
+2f_r
-\frac{k_{\mathrm{eff}}L}{2\pi}
\left(
\frac{1}{v_1}-\frac{1}{v_2}
\right)a,
\
\delta f_1+\delta f_2 &=
-\frac{k_{\mathrm{eff}}L}{2\pi}
\left(
\frac{1}{v_1}+\frac{1}{v_2}
\right)a.
\end{aligned}
\label{eq:MW-mea-1}
\end{equation}

Equation~(\ref{eq:MW-mea-1}) predicts that acceleration variations couple into the rotation channel through the difference between the two effective velocities. In contrast, rotation-to-acceleration coupling is theoretically absent because the rotational response is intrinsically velocity-independent. This behavior is directly verified by the cross-coupling measurements shown in the insets of Figs.~\ref{fig:inertial}(c,g). Furthermore, despite an approximately $5\%$ mismatch in the effective velocities of the two atomic beams, the individual envelope peak positions exhibit nearly identical rotational scale factors, as shown in Fig.~\ref{fig:inertial}(b). These results experimentally confirm that the rotational response is determined by the engineered velocity scaling rather than by the atomic velocity itself.

The velocity-independent rotational response is particularly important for practical quantum gyroscopes. This intrinsic robustness provides a direct route toward improved long-term scale-factor stability without requiring additional compensation or calibration procedures. More broadly, this result highlights a fundamental capability of white-light matter-wave interferometry: measurement characteristics can be deliberately engineered by tailoring the velocity dependence of different phase contributions. In the present work, matching the velocity scaling of the modulation-induced and rotation-induced phases yields a velocity-independent rotational response. More generally, velocity-domain engineering provides an additional degree of freedom for designing matter-wave interferometric observables and can be extended beyond inertial sensing to a broader class of quantum precision measurements.

\begin{figure}
\centering
\includegraphics[width=1\linewidth]{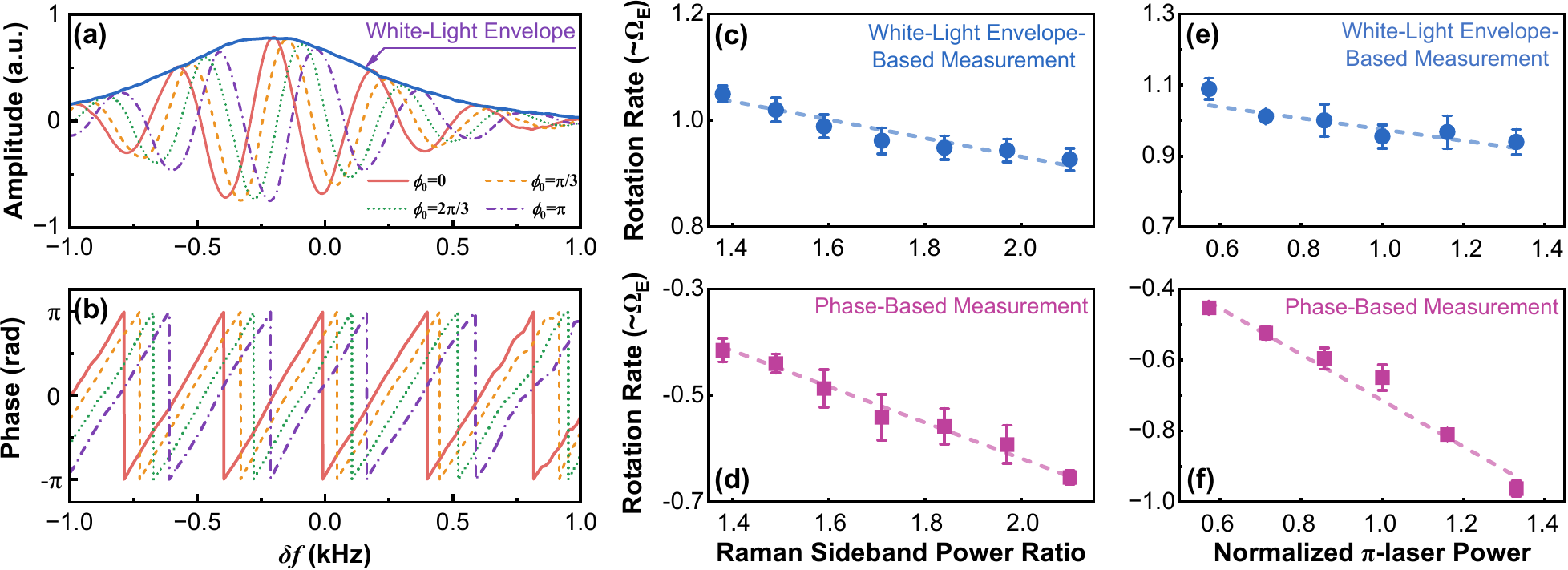}
\caption{Reduced sensitivity to interferometric phase bias in white-light matter-wave interferometry.
(a,b) Matter-wave interference fringes (a) and extracted phase (b) versus initial phase offset. The fringes shift within a common envelope [dashed line in (a)] under different operating phases.
(c,d) Sensitivity to light shifts for envelope-based measurement (c) and conventional phase-based measurement (d).
(e,f) Sensitivity to Raman laser power for envelope-based measurement (e) and conventional phase-based measurement (f).
}
\label{fig:absolute}
\end{figure}

\subsection{Feature III: Intrinsic Rejection of Phase Bias}

The third defining feature of white-light matter-wave interferometry is its intrinsic rejection of phase bias. Unlike conventional phase-based interferometry, in which velocity-independent phase offsets are directly mapped onto the measured signal, envelope localization is governed by the relative phase alignment across the broadband matter-wave spectrum. Consequently, phase contributions that are common to all velocity classes shift the underlying fringes but have little influence on the envelope peak position.

A representative phase bias in a Raman atom interferometer is the initial interferometric phase $\phi_0$, which is determined by the initial phases of the Raman fields. Because $\phi_0$ is independent of atomic velocity, it does not contribute to the stationary-phase condition in Eq.~(\ref{eq:partial}) and therefore does not shift the white-light envelope peak. This prediction is verified in Figs.~\ref{fig:absolute}(a,b), where $\phi_0$ is varied by modulating the phase of the RF drive applied to the Raman lasers. Although the underlying interference fringes and their extracted phases shift continuously with the imposed offset, they remain confined within a common interference envelope whose peak position is essentially unchanged. This behavior directly illustrates the distinction between conventional fringe-phase readout and white-light envelope localization: the former is sensitive to the absolute interferometric phase, whereas the latter is determined by the velocity-dependent phase-alignment condition.

We further investigate systematic perturbations associated with light shifts and Raman intensity variations. By tuning the intensity ratio of the Raman sidebands [Figs.~\ref{fig:absolute}(c,d)], the differential light shift is varied around its compensation point. The resulting sensitivity of the envelope-based rotation readout is reduced to approximately $50\%$ of that observed in conventional phase-based measurements. Similarly, variations in Raman laser power produce an approximately fourfold weaker response under envelope-based operation [Figs.~\ref{fig:absolute}(e,f)]. These results demonstrate that phase-related perturbations that predominantly act as common or weakly velocity-dependent offsets are substantially suppressed in the white-light interferometric observable.

For inertial sensing, this reduced sensitivity directly improves the robustness of the inferred rotation and acceleration against laser-phase drift, light-shift fluctuations, and power instability, thereby relaxing the requirements on long-term phase control and systematic calibration. More broadly, the result reveals a general property of white-light matter-wave interferometry: by defining the observable through spectral phase alignment rather than absolute fringe phase, the measurement can be made intrinsically insensitive to selected classes of phase perturbations. This provides an additional route for engineering robust matter-wave interferometric observables beyond inertial sensing.

\begin{figure}[t]
\centering
\includegraphics[width=0.65\linewidth]{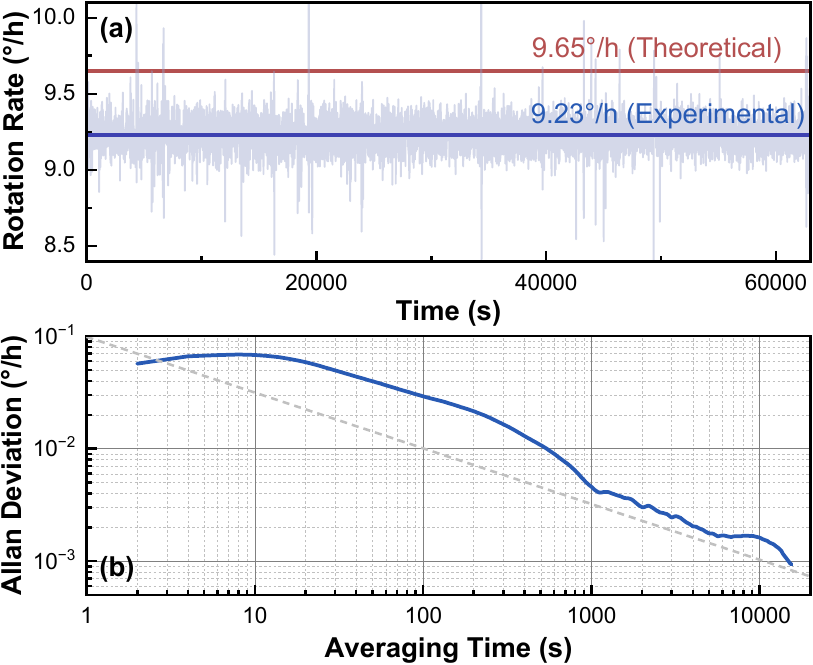}
\caption{Earth-rotation measurement using white-light matter-wave interferometry. (a) Continuous measurement of the Earth's rotation rate over 63,000 s. (b) Allan deviation of the measured rotation rate. The dashed line indicates the $1/\sqrt{\tau}$ dependence.}
\label{fig:earth}
\end{figure}

\subsection{Proof-of-Principle Demonstration: Earth-Rotation Measurement}

Having established the three defining features of white-light matter-wave interferometry, we finally examine whether the envelope-localized observable can support an absolute and long-term measurement of a real physical quantity. As a proof-of-principle demonstration, we measure the Earth's rotation, which provides a stable and independently known inertial reference without requiring an externally applied rotation input.

To suppress residual common-mode offsets and noise in the absolute readout, a $k$-vector reversal technique \cite{durfee2006long} is implemented by alternating between $+k_{\mathrm{eff}}$ and $-k_{\mathrm{eff}}$. The reversal changes the sign of the inertial response while allowing contributions from the modulation term $f_r$ and the envelope-fitting offset $e_{\mathrm{fit}}$ to be rejected through differential processing. Importantly, the resulting measurement is determined by the localized envelope positions and does not require precise reconstruction of the detailed envelope shape or prior knowledge of the atomic velocity distribution.

Continuous measurements are performed over $63,000~\mathrm{s}$ with a $k$-reversal period of $1~\mathrm{s}$, corresponding to an effective data rate of $0.5~\mathrm{Hz}$. The measured mean rotation rate is $9.23\,\mathrm{^\circ/h}$, in agreement with the expected projection of the Earth's rotation of $9.65\,\mathrm{^\circ/h}$ at the local latitude of $39.9^\circ$N in Beijing, China, with a relative error of approximately $4.3\%$. At an averaging time of $15,000~\mathrm{s}$, the Allan deviation reaches $9\times10^{-4}\,\mathrm{^\circ/h}$, corresponding to a relative stability of $93~\mathrm{ppm}$ with respect to the local Earth-rotation projection.

Rather than serving as a benchmark of gyroscope performance, this experiment provides a proof-of-principle validation of white-light matter-wave interferometry, demonstrating accurate, robust, and long-term absolute inertial measurement through envelope localization. The remaining deviation from the expected Earth-rotation projection may be associated with technical imperfections, including Raman pulse imperfections, wavefront distortions, beam misalignment, and leveling errors, which require further quantitative investigation.

Although demonstrated here through inertial sensing, this principle is not specific to rotation or acceleration. More fundamentally, the present work identifies spectral diversity as an additional degree of freedom for matter-wave interferometry: by engineering the spectral composition and its phase evolution, both the form and response of the resulting interferometric observable can be tailored. This perspective extends the role of broadband matter waves and establishes a connection to optical white-light interferometry. 
%White-light matter-wave interferometry may therefore provide a general framework for observable engineering in quantum precision measurement, opening opportunities for matter-wave sensors in which localization, response, and robustness are designed as intrinsic properties of the interference process.

\section{Conclusion}

We have experimentally realized white-light matter-wave interferometry, establishing the matter-wave counterpart of optical white-light interferometry. By exploiting the broadband velocity spectrum of an atomic beam, distinct matter-wave velocity classes are synthesized to form a localized interference envelope, directly paralleling the formation of the coherence envelope in optical white-light interferometry. Using inertial sensing as a proof-of-principle application, we demonstrate three defining features of this new interferometric regime: envelope localization beyond phase ambiguity, a velocity-engineered measurement response, and intrinsic rejection of phase bias. Accurate and long-term measurement of the Earth's rotation further demonstrates the capability of the envelope-localized observable for robust absolute measurement.

More fundamentally, white-light matter-wave interferometry introduces a new measurement mechanism by constructing a localized envelope observable through broadband matter-wave synthesis. This new observable enables phase ambiguity, measurement response, and systematic robustness to be addressed through the intrinsic design of the interferometric process itself. This work extends the paradigm of white-light interferometry from light waves to matter waves. Just as white-light interferometry established a distinct paradigm in optical metrology, its extension to matter waves provides a new framework for exploring and exploiting broadband matter-wave interference in quantum precision measurement. Future improvements in atomic source brightness, coherent control, and interferometer geometry are expected to extend the underlying concept beyond inertial sensing and open broader opportunities for observable engineering in matter-wave metrology.

%\bibliography{sample}

\end{document}